\documentclass[prl,twocolumn,superscriptaddress,aps,floatfix,showpacs,fleqn]{revtex4}
\usepackage{amsmath}
\usepackage{amsfonts}
\usepackage{amstext}
\usepackage{graphicx}
\usepackage{times}

\begin{document}

\title{An Ising-Glauber Spin Cluster Model for Temperature Dependent Magnetization Noise in SQUIDs}
\author{Amrit De}
\affiliation{Department of Physics and Astronomy, University of California - Riverside,CA 92521}
\date{\today}

\begin{abstract}
Clusters of interacting two-level-systems (TLS),likely due to $F^+$ centers at the metal-insulator interface, are shown to self consistently lead to $1/f^{\alpha }$ magnetization noise in SQUIDs. By introducing a correlation-function calculation method and without any a priori assumptions on the distribution of fluctuation rates, it is shown why the flux noise is only weakly temperature dependent with $\alpha\lesssim 1$, while the inductance noise has a huge temperature dependence seen in experiment, even though the mechanism producing both spectra is the same. Though both ferromagnetic- RKKY and short-range-interactions (SRI) lead to strong flux-inductance-noise cross-correlations seen in experiment, the flux noise varies a lot with temperature for SRI. Hence it is unlikely that the TLS's time reversal symmetry is broken by the same mechanism which mediates surface ferromagnetism in nanoparticles and thin films of the same insulator materials.
\end{abstract}



\maketitle

\def\ket#1{\left|#1\right\rangle}
\def\bra#1{\left\langle#1\right|}
\def\avg#1{\left\langle#1\right\rangle}



\label{sec.intro}

Superconducting quantum interference devices (SQUID) are of considerable interest for quantum information as they can replicate natural qubits, such
as electron and nuclear spins, using macroscopic devices. However the performance of superconducting qubits is severely impeded by the presence of $1/f$ magnetization noise which limits their quantum coherence. This type of noise was first observed in SQUIDs well over two decades ago\cite{Koch1983,Wellstood1987}, however its origins and many of its features remain unexplained. Recent activity in quantum computing has however revived interests to better understand this magnetization noise.\cite{Yoshihara2006,Bialczak2007}.

\begin{figure}[tbp]
\centering
\includegraphics[width=0.7\columnwidth]{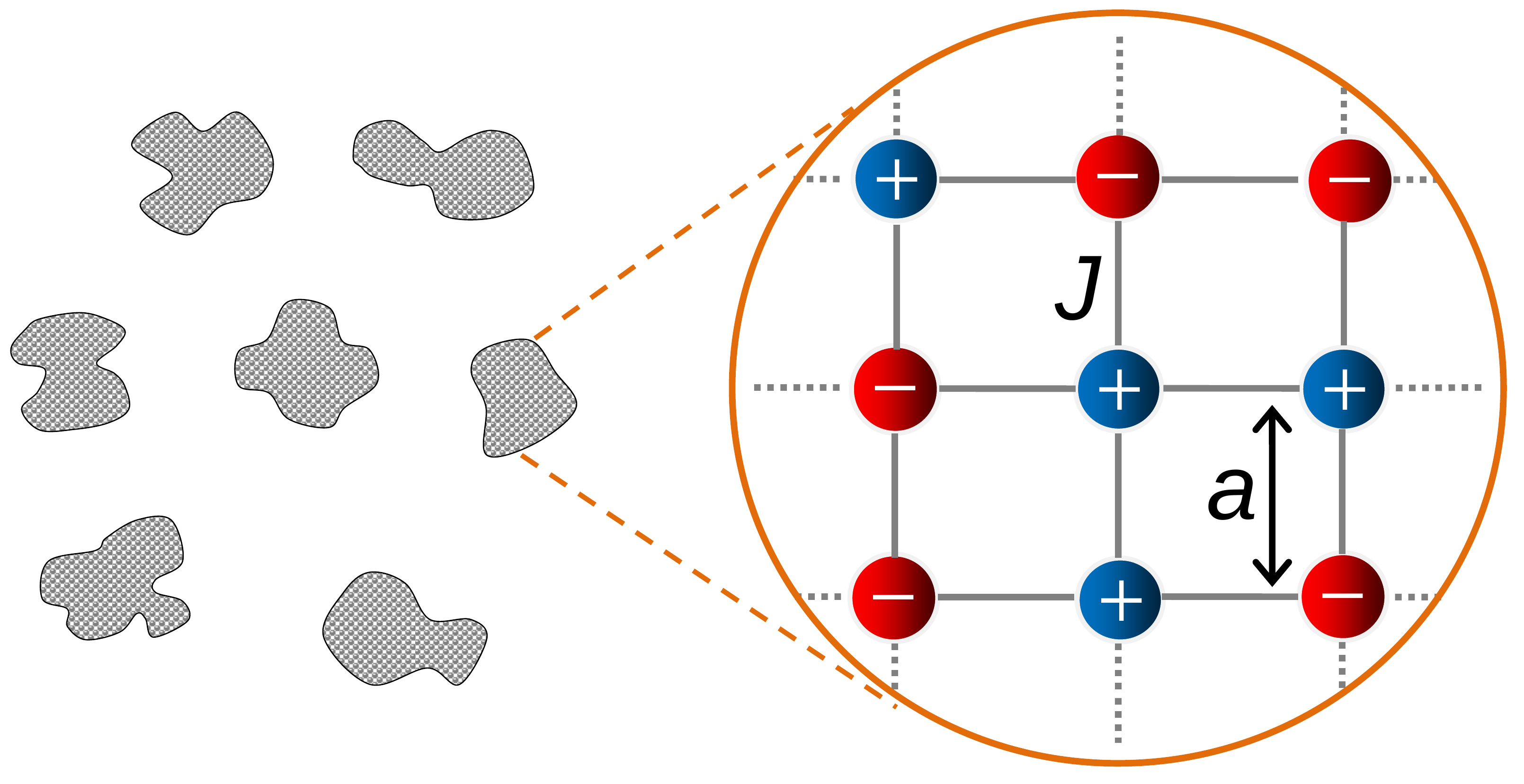}
\caption{Proposed $1/f$ noise model consisting of interacting spins that fluctuate within a cluster. The clusters are assumed to form due to random defects at the SQUID's metal-insulator interface and are sufficiently far apart so that only spins within a single cluster interact. Number of spins within a cluster and the lattice constant $a$ vary.
}
\label{fig:scheme}
\end{figure}

Magnetic noise in SQUIDs has several puzzling features. While the flux noise (the first spectrum) is also weakly dependent on temperature, the choice of the superconducting material and the SQUID's area\cite{Koch1983,Sendelbach2008,McDermott2009} -- the inductance noise (the second spectrum or the noise of the flux noise), surprisingly shows a strong temperature dependence. It decreases with increasing temperature and scales as $1/f^{\alpha }$\cite{Sendelbach2008} where the temperature dependent ($0<\alpha (T)<1$)\cite{Wellstood2011}.
The flux noise is also known to be only weakly dependent on geometry and the noise scales as $l/w$, in the limit $w/l<<1$ ($l$ is the length and $w$ is the width of the superconducting wire) \cite{Lanting2009}. This along with recent experiments \cite{Sendelbach2008} suggests that flux noise arises from unpaired surface spins which reside at the superconductor-insulator interface in thin-film SQUIDs. The estimated areal spin density from the
paramagnetic susceptibilities is about $5\times 10^{17}~m^{-2}$ \cite{Sendelbach2008,Bluhm2009}.

Experimental evidence also suggests that these surface spins
are strongly interacting and that there is a net spin polarization \cite{Sendelbach2008} as the $1/f$ inductance noise is highly
correlated with the usual $1/f$ flux noise. This cross-correlation is
inversely proportional to the temperature and is about the order of unity
roughly below 100mK. Since inductance is even under time inversion and flux is odd, their three-point cross-correlation function must vanish unless time reversal symmetry is broken, which indicates the appearance of long range magnetic order. As this further implies that the mechanism producing both the flux- and inductance noise is the same, it is not clear on why only the associated spectrum (inductance noise) should have a large temperature dependence\cite{Sendelbach2008}.


Usually $1/f$ noise is associated with the onset of a spin-glass phase as is believed to account for the many anomalous
properties of spin-glasses at low temperatures\cite{Weissman1993,Wu2005}. However recent Monte-Carlo simulations by Chen and Yu \cite{Chen2010} have ruled out this out to explain magnetization noise in SQUIDs. They considered an Ising spin glass with random nearest neighbor interactions. Though their model reproduced qualitative features of the inductance noise, it did not show cross-correlations between inductance and flux noise. This is expected as spin glasses preserve time reversal symmetry.

The microscopic origin of the magnetization noise is also not known. Phenomenologically, the noise arises from spins bereave like randomly fluctuating two level systems(TLS). Choi {\it et al}\cite{Choi2009} explained this in terms of metal induced gap states that arise due to the potential disorder at the metal-insulator interface. Another model involved unpaired and non-interacting electrons randomly hopping between traps with different spin orientations and a $1/f$ distribution of trap energies\cite{Koch2007}. Quite often hopping conductivity models are used for $1/f$ noise in solid state systems\cite{Shtengel2003,Shklovskii2003,Burin2006}. Dangling bond states near the Fermi energy in disordered SiO$_{2}$
\cite{deSousa2007} and fractal spin structures with tunneling interactions\cite{Kechedzhi2011} have also been suggested for the SQUID noise.


Typically, the dc-SQUIDs consitute of an amorphous Al$_{2}$O$_{3}$ insulating layer deposited on the surface of a metal (commonly Nb\cite{Sendelbach2008} or Al\cite{Bialczak2007}). The Al$_{2}$O$_{3}$ is likely to cluster on the surface before filling in and forming a homogeneous layer due to its higher binding energy which could lead to the Volmer-Weber growth mode. The lattice mismatch between the insulator and the metal could also lead to the formation of clusters. Near the metal surface, the clusters can host a number of point defects in the form of O vacancies that can capture one electron (F$^+$-center) or two (F-center).

In a related development, a few years ago surface ferromagnetism (SFM) was reported in thin-films and nanoparticles of a number of otherwise insulating metallic oxides\cite{Sundaresan_prb2006} (including Al$_2$O$_3$) where the materials were not doped with any magnetic impurities. Further recent investigations attribute this room temperature SFM  in Al$_2$O$_3$ nanoparticles\cite{Yang_JPC2011} to F$^+$-centers where it was found that amorphous Al$_2$O$_3$ is more likely to host the number of F$^+$-centers to cross the magnetic percolation threshold than the crystalline variant.

The origin of SFM in these otherwise non-magnetic metal oxides is itself somewhat controversial\cite{Keating2009}. Some of the suggested mechanisms include exchange coupling from F$^+$ center induced impurity bands\cite{Venkatesan2004}, F$^+$ center mediated superexchange\cite{Han_prb2009}, and spin-triplets at the F-center\cite{Chang_PRB2012}.
In addition to this, in the SQUID geometry because of the proximity to the metal, these local magnetic moments can spin polarize the metal's conduction band electrons which can lead to an RKKY type long range interaction mechanism, which was first pointed out by Faoro and Ioffe\cite{Faoro2008}. This can likely lead to competing interaction mechanisms.

In this paper, in order to see if the interaction mechanism induces any distinguishing features in the noise spectrum, both ferromagnetic RKKY and ferromagnetic nearest neighbor interaction(NNI) mechanisms are considered for the spin cluster model. While both mechanisms give rise to $1/f$ noise self consistently as shown here, for NNIs the flux noise varies far more with temperature.

A new method is introduced in this paper to obtain any arbitrary $n-$point correlation function. As a result various subsequent spectral functions for the interacting Ising-Glauber spin model can be systematically obtained, both analytically and numerically. The magnetization noise calculations are carried out using this method and the spin cluster model shown in fig.\ref{fig:scheme}. For the sake of self consistency, the usual heuristic assumption on the $1/\gamma$ distribution of switching rates required for $1/f$ noise\cite{Kogan.book}) is avoided. Instead it is shown that at low temperatures, the $1/f$ spectrum arises naturally in the model from the spin-spin interactions and a uniform distribution of cluster sizes. All experimentally observed features such as the lack of temperature dependence of the flux noise, the previously unexplained considerable temperature dependence of the inductance noise (or the associated spectrum) and the flux-inductance noise cross-spectrum are shown and explained here.

\label{sec.1}
\emph{The Model and the Method:} A schematic of our model is shown in figure\ref{fig:scheme} where each cluster comprises of an interacting 2D spin-lattice. To model the randomness of the surface defects, randomly varying lattice constants are considered. Individual clusters are assumed to be sufficiently far apart so that there are no interactions across the clusters but there could be an effective mean field. For the infinite range Ising-Glauber model, all $N$ spins within a single cluster interact with every other via an RKKY type mechanism.

The overall temporal evolution for $N$ interacting spins is governed by the master equation $\dot{\mathbf{W}}(t)=\mathbf{VW\mathrm{\mathit{(t)}}}$ \cite{VanKampen.book}, where $\mathbf{V}$ is a matrix of transition rates (such that the sum of each of its columns is zero) and $\mathbf{W}$ is the
flipping probability matrix for the spins. Each spin's random fluctuation is a temperature (or interaction) driven process governed by Glauber dynamics. This is a Markov process where the new spin distribution depends only on the current spin configuration and that the new and old spin configurations agree everywhere except at a single site. Overall the non-equlibrium spin dynamics for a system of correlated spins can be treated this way\cite{Ozeki1997}. The conditional probability for a single spin to flip is determined by the Boltzmann factor and the matrix-elements of $\mathbf{V}$ are
\begin{equation}
\small{
{\mathbf{V}}(\mathbf{s}\rightarrow \mathbf{s}^{\prime })=\left\{
\begin{array}{ll}
\displaystyle\frac{\gamma e^{-\beta H(\mathbf{s^{\prime }})}}{e^{-\beta H(%
\mathbf{s})}+e^{-\beta H(\mathbf{s^{\prime }})}} &
\mbox{for
$\mathbf{s}\neq\mathbf{s}'$ and} \\
& \mbox{$\displaystyle\sum_i(1-s_is_i')=2$} \\
-\displaystyle\sum_{\mathbf{s}\neq \mathbf{s}^{\prime }}V(\mathbf{s}%
\rightarrow \mathbf{s}^{\prime }) & \mbox{for $\mathbf{s}=\mathbf{s}'$}%
\end{array}%
\right.}   \label{Glaub}
\end{equation}%
Here, $\mathbf{s}^{\prime }$($\mathbf{s}$) is a vector that denotes the present(earlier) spin configuration and $\gamma$ is the flipping rate of a spin (all $\gamma=1$ in this paper). The non negative off-diagonal matrix elements in Eq.\ref{Glaub} satisfy the detailed balance condition and the
diagonal terms are the just negative sum of the off-diagonal column elements so that the column's zero sum ensures the conservation of probability. To obtain the random temporal dynamics of an individual spin cluster, the $2^N$ dimensional $\mathbf{W}=\exp(-{\bf V}t)$ matrix has to be evaluated. The eigenvalues of $\mathbf{V}$ are either zero, which corresponds to the equilibrium distribution, or are real and negative, which also eventually tend to the equilibrium distribution as $t\rightarrow \infty $ \cite{VanKampen.book}. This method is also provides the quasi-Hamiltonian open quantum systems methods\cite{Joynt2011,Zhou2010,De2013pra} with a connection to the underlaying noise microscopics.

The general system Hamiltonian is
\begin{equation}
H(\mathbf{s})=-\frac{1}{2}\sum_{i,j}J_{i,j}s_{i}s_{j}-B\sum_{i}s_{i}
\label{H_ising}
\end{equation}%
where $B$ is the magnetic field(which is set to zero here) and $J_{i,j}$ is the spin-spin interaction
between the $i^{th}$ and $j^{th}$ Ising spins. Now, for $N$ interacting spins, the $n^{th}$ order auto- or cross-correlation function can be calculated as follows
\begin{eqnarray}
\langle s_{i}(t_{1})s_{j}(t_{2})...s_{\kappa}(t_{n})\rangle = \langle {\mathbf{ f}}%
|\sigma_z^{(\kappa)}\mathbf{W}(t_n) ... \sigma_z^{(i)}\mathbf{W}(t_1)|\mathbf{ i}\rangle  \label{gencorr}
\end{eqnarray}
where the spin indices $\{i,j...\kappa\}\in\{1,2,...N\}$, $|\mathbf{i}\rangle $=$|\mathbf{f}\rangle$ are the initial and final state vectors that correspond to the equilibrium distribution (\emph{i.e.,}$\mathbf{W}\ket{\mathbf{i}}=\ket{\mathbf{i}}$).
It is implied that $\sigma_z^{(\kappa)} = \underset{1}{\sigma_o}\otimes\underset{2}{\sigma_o}...\underset{\kappa-1}{\sigma_o}\otimes\underset{\kappa}{\sigma_z}\otimes ... \underset{N}{\sigma_o}$
where $\sigma_z$ is the z-Pauli matrix and $\sigma_o$ is the identity.
For just two spins, if all $\gamma_i=1$ (which is subsequently followed for all calculations) the two-point correlation functions are
\begin{equation}
\langle{s_i(0)s_j(t)}\rangle= e^{-2\Gamma_-|t|} + (2\delta_{ij}-1)\exp(4\beta J)e^{-2\Gamma_+|t|}
\label{Tpt-G}
\end{equation}
where $\Gamma_{\pm}=[1+\exp(\pm2\beta J)]^{-1}$. Whereas if $\gamma_i$ is retained then $\displaystyle\lim_{T\rightarrow0}$ $\langle{s_is_j}\rangle=\delta_{ij}e^{-2\gamma_i|t|}$ is obtained from this model.


Within a single cluster, the spins interact via an oscillatory RKKY-like form with a ferromagnetic $J_o$,
$J_{i,j}=J_{o}{[k_FR_{i,j}\cos (k_FR_{i,j})-\sin (k_FR_{i,j})]}/{(k_FR_{i,j})^{4}}$
where $R_{i,j}$ is the separation between two spins (on a lattice of lattice constant $a$), $k_F$ is a Fermi wavevector type parameter. For the calculations here $J_o=9\times 10^{10}~Hz/\hbar$ is taken as a fitting parameter independent of $k_F$.



\begin{figure}
\centering
\includegraphics[width=1\columnwidth]{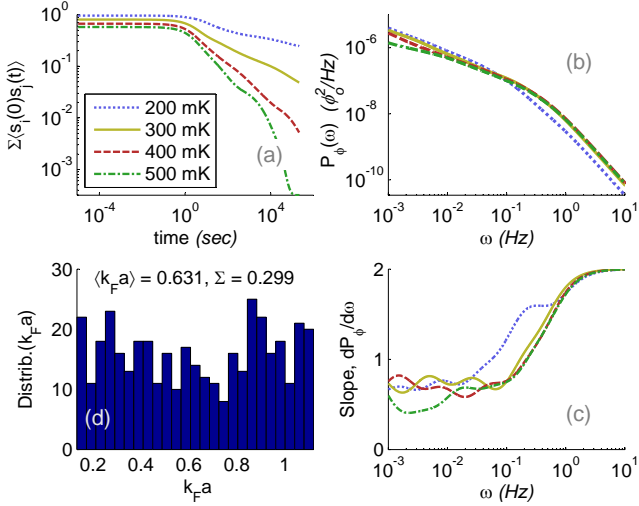}
\caption{ \textbf{(a)} Temperature dependent net correlation function for 400 spin clusters, each
with 6-9 spins with ferromagnetic ($+|J_o|$) RKKY interactions. \textbf{(b)} Corresponding flux noise power-spectrum showing $1/f^{\alpha}$ noise below $\sim0.1$ Hz and \textbf{(c)} its respective slope ($\alpha$). Note that $\alpha\lesssim 1$ below about $0.1$ Hz \textbf{(d)} Distribution of $k_Fa$ for each cluster, with mean $\langle{k_Fa}\rangle$ and standard deviation ($\Sigma$) as indicated.}
\label{fig:P1}
\end{figure}

\emph{Flux Noise:}
Due to the high estimated areal spin density, the coherent magnetization of the spins strongly flux couples to the SQUID. The fluctuation-dissipation theorem relates the the magnetization noise spectrum to the imaginary part of the susceptibility\cite{Reim1986,Vitale1989,McDermott2009}. If all the surface spins couple to the SQUID equally, the flux noise for the $\ell^{th}$ spin-cluster is the Fourier transform of the sum of all two-point spin correlation functions
\begin{equation}
P_{\phi }^{(\ell)}(\omega )=2\mu _{o}^{2}\mu _{B}^{2}\frac{\rho }{\pi }%
\frac{R}{r}\displaystyle\int_{0}^{\infty }\displaystyle\sum_{i,j=1}^{N}\langle {s_{i}(0)s_{j}(t)}\rangle e^{\imath\omega t}dt~~
\end{equation}%
where $R$ is the radius of the loop, $r$ is the radius of the wire, $R/r=10$\cite{McDermott2009} and $\rho$ is the surface spin density. All possible combinations of two-point autocorrelation ($i=j$) and cross-correlation ($i\neq j$) functions are explicitly calculated within a cluster using Eq.\ref{gencorr}. Since the clusters are assumed to be sufficiently far apart and noninteracting, $\langle{s_{i}(0)s_{j}(t)}\rangle$ is calculated individually for each cluster. The total flux noise power spectrum for the SQUID is $P_{\phi }(\omega )=\sum_{\ell}{P_{\phi}^{(\ell)}(\omega )}$.

The temperature dependent net $P_{\phi }(\omega )$, and the respective noise-slopes are shown in Figs.\ref{fig:P1}(a)-(c), where 400 spin clusters were considered with 6-9 spins each. Each cluster is assigned a random $k_Fa$ and a uniform distribution of $k_Fa$ is considered (see Fig.\ref{fig:P1}-(a)). Note that from the estimated areal spin density of unpaired surface spins of $\rho\sim5\times10^{17}~m^{-2}$, one can estimate $k_F$ from which we can infer what the average spin separation $\langle{a}\rangle$ is.

The $1/f^{\alpha}$ noise spectrum (indicated by the slope $\sim 1$) is shown in Fig.\ref{fig:P1} (b) and (c), at for the intermediate range of frequencies. At high frequencies, the \emph{log} noise spectra shows a slope of $2$ which corresponds to the Lotrentzian tail of the noise power. While considering various cases, it was found that the upper and lower cutoff frequencies for the $1/f$ type noise depended on the distribution of $k_{F}a$ and the interaction strength (as evident from the temperature dependence). In the absence of interactions the noise spectrum reduces to a simple Lorentzian. And eventually for all temperatures, $\alpha \rightarrow 0$ (as determined by the interaction strength) at very low frequencies which corresponds to Gaussian noise.

These calculations show that the noise power is only weakly dependent on temperature which is in agreement with experiment\cite{Sendelbach2008}. Note that in the early experiments of REF.[\onlinecite{Wellstood1987}], $P_{\phi }$ was not necessarily independent of temperature under \textit{all} circumstances. While there was no temperature dependence for the flux noise bellow 1K, there was a strong low temperature dependence for certain parameters/materials -- for example, for PbIn/Nb and Pb/Nb \cite{Wellstood1987}. Quite strikingly, for the same set of materials (PbIn/Nb for the SQUID's loop/electrode) and depending on the construction, the flux noise can be either completely independent of temperature or inversely proportional to it or even oscillate with temperature. Such conflicting temperature dependencies of $1/f$ noise are known to exist for glassy systems\cite{Massey1997,McCammon2002}.

\begin{figure}
\centering
\includegraphics[width=1\columnwidth]{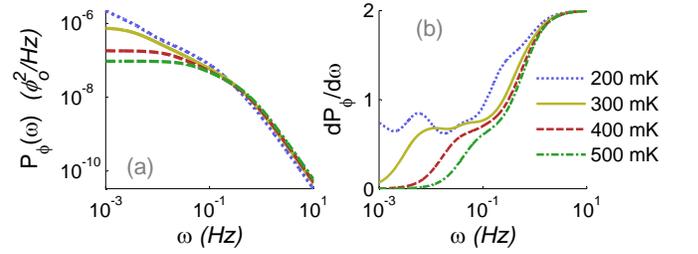}
\caption{ \textbf{(a)} Sample flux noise power-spectrum $P_\phi$ and \textbf{(b)} its slope for spin clusters with nearest neighbor ferromagnetic interactions that vary randomly for each cluster. }
\label{fig:P1ii}
\end{figure}
Next consider short range ferromagnetic NNI $\mathcal{J}$ that randomly vary for each cluster. Now $1/f$ noise (see fig.\ref{fig:P1ii}) can only be obtained in the spin-cluster model if the NNIs have a $1/{\mathcal{J}}$ type distribution. If the RKKY $J_{i,j}$ is expanded for small $k_F$, then $J_{i,j}\propto{1/R_{i,j}}$ -- hence a uniform distribution of $R_{i,j}$ results in a $1/J_{i,j}$ distribution of interaction strengths at a certain crossover length $1/k_F$. Although this ferromagnetic NNI model also self consistently produces $1/f$ noise, the flux noise's variations with temperature are quite large (see Fig.\ref{fig:P1ii}) and is $1/f$-like only over a short range (compare to fig.\ref{fig:P1}). In view of the experiments\cite{Sendelbach2008}, this likely rules out NNIs being the dominant mechanism even though they are the thought to be the cause of SFM in insulating metal oxide nanoparticles\cite{Venkatesan2004,Sunderasan2009,Han_prb2009}.

\emph{Inductance Noise:}
In the experiments of Ref.\onlinecite{Sendelbach2008}, the temperature
dependent inductance noise was measured for temperatures bellow $2K$ where the inductance noise($P_L$) was mostly dominated by the imaginary
part of the susceptibility and varied considerably with temperature. $P_L$ is the associated noise spectrum or the second spectrum -- which is a quantitative measure of the spectral wandering of the first spectrum and is interpreted as the
noise of the noise \cite{Nguyen2001}. The first spectrum (flux noise) is related to the imaginary part of the susceptibility via the
fluctuation-dissipation theorem $P(\omega )\approx 2{k_{B}T}\chi ^{\prime \prime }/{\omega}$. Assuming all spins couple to the SQUID equally \cite{Zhou2010}, the imaginary part of the inductance then relates to the spin susceptibility within a layer of thickness $d=\rho /\tilde{n}$ on the surface as $L^{\prime \prime }=\mu _{o}d\frac{R}{r}\chi ^{\prime \prime }$, therefore
\begin{eqnarray}
P_{L}(\omega ) 
&=&\left( \mu _{o}d\frac{R}{r}\right) ^{2}\int_{0}^{\infty }\langle \chi
(0)\chi (t)\rangle e^{\imath\omega t}dt  \label{PL}
\end{eqnarray}%
and from the fluctuation dissipation theorem,
\begin{equation}
\chi ^{\prime \prime }(\omega )=2\frac{\tilde{n}\mu _{o}\mu _{B}^{2}\omega }{%
k_{B}T}\displaystyle\sum_{i,j}\displaystyle\int_{0}^{\infty }\langle
s_{i}(0)s_{j}(t)\rangle e^{\imath\omega t}dt.
\end{equation}%
It is argued here that the sum of all two-point correlation functions for the system of interacting spins can always be expressed as $\sum\langle{s_{i}(0)s_{j}(t)}\rangle=\sum {C_{\nu }e^{-2\Gamma _{\nu }t}}$, which is shown analytically for two spins (see Eq.\ref{Tpt-G}) and systematically verified for more numerically. Hence
\begin{equation}
\chi ^{\prime \prime }(\omega )=2\tilde{n}\mu _{o}\mu _{B}^{2}\frac{\omega }{%
k_{B}T}\sum_{\nu }\frac{C_{\nu }\Gamma _{\nu }}{\Gamma _{\nu
}^{2}+\omega ^{2}}
\end{equation}%
and the real part is
\begin{eqnarray}
\chi^{\prime}(\omega)=\frac{2}{\pi }{\mathcal{P}}\displaystyle\int_{0}^{\infty }\frac{\chi ^{\prime\prime }(\omega^{\prime} )}{\omega ^{\prime 2}-\omega ^{2}}d\omega ^{\prime }=\frac{2\tilde{n}\mu_{o}\mu_{B}^{2}}{k_{B}T}\sum_{\nu }\frac{C_{\nu }\Gamma _{\nu }^{2}}{\Gamma _{\nu }^{2}+\omega ^{2}}~
\end{eqnarray}%
where, $\mathcal{P}$ is Cauchy's principal value. Hence from the total susceptibility $\chi(\omega)=\chi^{\prime}(\omega)+i\chi ^{\prime\prime}(\omega)$,
\begin{eqnarray}
\chi (t)=\int_{0}^{\infty }\chi(\omega)e^{\imath\omega t}d\omega
=\frac{2\tilde{n}\mu _{o}\mu _{B}^{2}}{k_{B}T}\sum_{i,j}\langle
s_{i}(0)s_{j}(t)\rangle.~~~
\label{Xt}
\end{eqnarray}
The inductance noise can then be explicitly expressed in terms of the
spectral density of the dynamical four-point noise correlation functions\cite{Kogan.book},
\begin{eqnarray}
P_L(\omega) = \left(2\rho\frac{\mu_o^2\mu_B^2}{k_BT} \frac{R}{r}%
\right)^2 \displaystyle\iint\limits_{\omega_{a}}^{~~~~%
\omega_{b}}S^{[2]}(\omega,\omega_1,\omega_2)d\omega_1 d\omega_2~~~~~~
\label{PL2}
\end{eqnarray}
where
$S^{[2]}(\omega,\omega_1,\omega_2)=\iiint\limits_0^{~~~~\infty}\sum\langle s_i(t_1)s_j(t_2)s_k(t_3)s_l(t_4)\rangle$
$e^{\imath(\omega_1-\omega)\tau^{\prime }}e^{\imath(\omega_2+\omega)\tau^{\prime\prime }}e^{\imath\omega\tau}d\tau^{\prime} d\tau^{\prime\prime} d\tau$
here $\tau^{\prime }=t_2-t_1$, $\tau^{\prime \prime }=t_4-t_3$ and $\tau=t_4+t_3-t_2-t_1$. $\Delta\omega=\omega_{b}-\omega_{a}$ is the
bandwidth within which the second spectrum is observed
%
\begin{figure}[tbp]
\centering
\includegraphics[width=1\columnwidth]{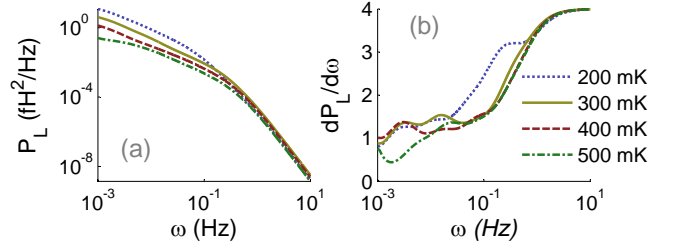}
\caption{ \textbf{(a)} Power-spectrum of the inductance noise, $P_L$ (associated noise of $P_\phi$ of fig.\ref{fig:P1}) and \textbf{(b)} its respective slope ($\protect\alpha$) for the spin cluster model with ferromagnetic $(+|J_o|)$ RKKY interactions.
The integrated $\avg{\alpha}$ between 0.001-0.05 Hz is \{1.569,1.415,1.247,1.242\} for the respective temperatures in ascending order.
}
\label{fig:P2}
\end{figure}

The temperature-dependent inductance noise spectrum and its slope is shown in fig.\ref {fig:P2}, where $\Delta\omega$ is set to cover the full spectrum. The noise power spectrum now shows $1/f^{\alpha }$ behavior at intermediate frequencies where the average integrated $\alpha$ between $0.001-0.05~Hz$ varies from $\sim 1.57$ (at $200~mK$) to $\sim1.24$ (at $500~mK$). And for even higher temperatures $\alpha\rightarrow 0$. The $\alpha=4$ at high frequencies is due to the square of the Lorentzian tail while at the lowest frequencies $\alpha$ eventually rolls over to zero. This onset of gaussian noise type behavior at low frequencies is again determined by the temperature. Overall the temperature dependent $1/f$ inductance noise behavior for the spin cluster model used here agrees very well with experiment\cite{Sendelbach2008,Wellstood2011}.

\begin{figure}[tbp]
\centering
\includegraphics[width=1\columnwidth]{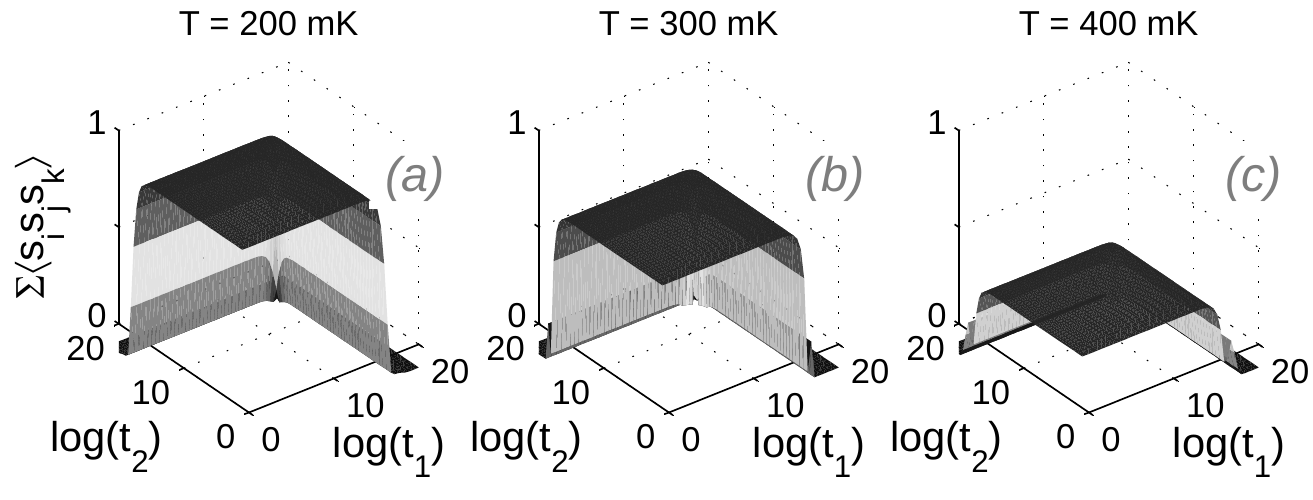}
\caption{Normalized sum of all three-point correlation functions, {\protect\small $%
\sum\langle{s_i(0)s_j(t_1)s_k(t_2)}\rangle/N^3$}, indicating flux-inductance-noise crosscorrelation for a single cluster of $N=10$
spins with ferromagnetic ($+|J_o|$) RKKY interactions at temperatures of (\textbf{a}) $T=200~mK$ (%
\textbf{b}) $T=300~mK$ and (\textbf{c}) $T=400~mK$.}
\label{fig_3ptC}
\end{figure}

\emph {Flux-Inductance-Noise Cross-Correlation:} Finally, the SQUID's surface spins spins show a net polarization in the experiments\cite{Sendelbach2008} as the $1/f$ inductance noise was found to be highly correlated with the $1/f$ flux noise. The following expression gives the flux- and inductance noise cross power spectrum
\begin{eqnarray}
P_{L\phi}(\omega )=\frac{1}{k_{B}T}\left( 2\rho \mu _{o}^{2}\mu
_{B}^{2}\frac{R}{r}\right)^{\frac{3}{2}}\times~~~~~~~~~~~~~\\\nonumber
\int\limits_{\omega _{a}}^{\omega _{b}}\iint\limits_{0}^{~~~~\infty }
\displaystyle\sum_{i,j,k}\langle s_{i}(t_{1})s_{j}(t_{2})s_{k}(t_{3})\rangle
e^{\imath\omega _{-}\tau }e^{\imath\omega _{+}\tau ^{\prime }}d\tau d\tau ^{\prime
}d\omega^{\prime}
\label{Pw3pt}
\end{eqnarray}%
where, $\tau =t_{2}-t_{1}$, $\tau ^{\prime }=t_{3}-t_{2}$, $\omega _{\pm}=\omega \pm \omega ^{\prime }$ and $\omega _{b}-\omega _{a}$ defines the
bandwidth. In the experiments $P_{L\phi}$ was found to be inversely proportional to temperature and about $\sim1$ at temperatures roughly below $100mK$. Now $P_{L\phi}$ depends on the sum of all three-point auto- and cross-correlation functions(TPCF). As inductance is even under time inversion and magnetic flux is odd, the flux-inductance-TPCF can only be nonzero if time reversal symmetry is broken -- indicating the appearance of long range magnetic order. This indicates that the interactions must be ferromagnetic. To show this, the TPCF (for all possible spin combinations) is calculated by for a single cluster of $10$ spins with ferromagnetic RKKY interactions, which gives $\sum\langle{s_{i}s_{j}s_{k}}\rangle_{max}\sim 1$ at low temperatures and keeps decreasing as the temperature is increased (see Fig.\ref{fig_3ptC}). This is in excellent agreement with experiment and also verifies that the mechanism mechanism produces both the flux noise and inductance noise. Whereas for antiferromagnetic RKKY interactions, $\sum\langle {s_{i}s_{j}s_{k}}\rangle_{max}\sim 0$.


%
%

\emph{Summary:}
Overall, previously unexplained experimentally observed features of the temperature dependent $1/f^\alpha$ magnetization noise in SQUIDs is explained by an Ising-Glauber spin-cluster model. The inductance noise is inherently $T^{-2}$ dependent while the flux noise is not. A general method is introduced for obtaining $n-$point correlation functions and various spectral functions subsequently. Explicit flux-inductance cross-correlation function calculations suggest that ferromagnetic RKKY interactions between $F^+$ centers at the metal-insulator interface are the most likely cause of the observed long range magnetic ordering of the TLSs.



I wish to thank Robert Joynt for a number of invaluable discussions and for his comments on this work. I would like to thank Robert McDermott for carefully explaining the experiments. And I would like to thank Leonid Pryadko for his support. This work was done partially under DARPA-QuEst Grant No. MSN118850 and presently with the support of U.S. Army Research Office Grant No.~W911NF-11-1-0027 and NSF Grant No.~1018935.

\bibliographystyle{apsrev}

\end{document}